\documentclass[preprint]{aastex}

\def\kms{km~s$^{-1}$}
\shorttitle{CTA~21: Discovery of H{\sc i} absorption}
\shortauthors{Salter et al.}

\begin{document}

\title{The Discovery of Host Galaxy H{\sc i} Absorption in CTA~21}

\author{C.J. Salter,$^1$ D.J. Saikia,$^{2,3}$
R. Minchin,$^1$ T. Ghosh$^1$ and Y. Chandola$^2$}
\affil{$^1$ Arecibo Observatory, NAIC, HC3 Box 53995, Arecibo, Puerto Rico, PR 00612, USA}
\affil{$^2$ National Centre for Radio Astrophysics, TIFR, Post Bag 3, Pune 411 007, India}
\affil{$^3$ ICRAR, University of Western Australia, Crawley, WA 6009, Australia}

\begin{abstract}
We report the discovery of H{\sc i} 21-cm absorption towards the
well-studied GHz Peaked-Spectrum source CTA~21 (4C~16.09) using
the Arecibo Telescope on 2009 September 20 and 21. Recently, the
frequency band between 700 and 800 MHz was temporarily opened up to
radio astronomy when US TV stations were mandated to switch from
analog to digital transmissions, with new frequency allocations.  The
redshifted H{\sc i} frequency  for CTA~21 falls within this band.
CTA~21 has a complex radio structure on a range of scales.  The
innermost prominent components are separated by $\sim$12 mas while weak
diffuse emission extends for up to $\sim$300 mas.  The H{\sc i}
absorption profile that we find has two main components, one narrow,
the other wider and blue-shifted. The total H{\sc i} column
density is 7.9$\times$10$^{20}$ cm$^{-2}$, assuming a covering
factor of unity and a spin temperature of 100 K.  This H{\sc i}
absorption confirms the recently determined optical redshift of this
faint galaxy of z$\sim$0.907.  We discuss this new detection in the
light of H{\sc i} absorption studies towards compact radio sources, and
also the possibility that CTA~21 may be exhibiting multiple cycles of
nuclear activity.  This new detection in CTA~21 is
consistent with a strong trend for detection of H{\sc i} absorption 
in radio galaxies with evidence of episodic nuclear/jet activity.   
\end{abstract}

\keywords {galaxies: active --- galaxies: nuclei --- galaxies:
individual (CTA~21) --- radio lines: galaxies}

\section{Introduction}
In the widely accepted model of active galactic nuclei (AGN), the
source of energy is the gravitational potential energy of the material
being accreted by a supermassive black hole via an accretion disk (e.g.
Krolik 1999). This nuclear region is surrounded by a torus consisting
of ionized, atomic and molecular gas components (e.g. Urry \&
Padovani 1995 and references therein). Studying the kinematics and
distribution of the gaseous components in the circumnuclear region is
important for understanding a number of aspects, such as the fueling
of the AGN activity, the anisotropy of the radiation field and thereby
testing the unified schemes for active galaxies, interaction of the
jets with the external gas clouds and probing star formation in the
central regions of an AGN. At radio wavelengths, the ionized component
may be investigated via polarization observations of compact radio
structure in the nuclear regions of AGNs (e.g.
Saikia \& Salter 1988; Mantovani et al. 1994; Udomprasert et al.  1997;
Junor et al.  1999; Saikia \& Gupta 2003; Rossetti et al.  2009).
These include the compact radio cores, the nuclear radio jets, and the
compact steep spectrum (CSS) and Gigahertz peaked spectrum (GPS)
sources.  CSS sources are defined as having a projected linear
size $<$15 kpc (H$_o$ = 71 km s$^{-1}$ Mpc$^{-1}$, $\Omega_{\rm m}$ =
0.27, and $\Omega_{\Lambda}$ = 0.73), and  a steep high-frequency
radio spectrum ($\alpha > 0.5$, where S$(\nu)\propto\nu^{-\alpha}$).
GPS sources have spectra which turn over at, or above, 1 GHz, and
are more compact than CSS objects  whose spectra could turn over below
1~GHz.  It is believed that GPS sources evolve into CSS objects, which
later evolve to form the larger radio galaxies and quasars (Fanti et
al.  1995; Readhead et al.  1996; O'Dea 1998; Snellen et al. 2000).

An important way of investigating atomic gas on subgalactic scales is
via H{\sc i} absorption towards the compact components of CSS and GPS
sources or the radio nuclei of larger objects (e.g. van Gorkom et al.
1989; Conway \& Blanco 1995; Peck et al. 2000; Pihlstr\"om et al.
2003; Vermeulen et al. 2003; Gupta et al. 2006; Morganti et
al. 2009 and references therein).  These studies have shown that
approximately 50\% of GPS objects exhibit H{\sc i} absorption
compared with about 35\% for  CSS sources (cf. Gupta et al.  2006).
The H{\sc i} column density also exhibits an anticorrelation with
source size (Pihlstr\"om et al. 2003; Vermeulen et al. 2003; Gupta et
al. 2006). The H{\sc i} spectra exhibit a variety of line profiles with
substantial red and blue shifts from the systemic velocities
(Vermeulen et al. 2003; Gupta et al.  2006), suggesting that the atomic
gas possesses complex motions, and may be out-flowing or in-falling,
interacting with the jets, or rotating around the nucleus.

Another interesting class of AGN are those which exhibit signs of
episodic activity. In radio-loud AGN these can be seen as two or more
pairs of radio lobes on opposite sides of the active nucleus, or as
young radio lobes embedded in diffuse emission from earlier cycles
of activity. The objects with two or more pairs of distinct lobes have
been designated double-double radio galaxies or DDRGs (e.g.
Schoenmakers et al. 2000; Saikia et al. 2006).  Saikia et al. (2007)
reported the detection of H{\sc i} absorption towards the central
region of the DDRG J1247+6723, and suggested from available information
that there might be a strong correlation between the detection of H{\sc
i} in absorption and the occurrence of rejuvenated radio activity. The
detection of H{\sc i} absorption in the rejuvenated radio galaxy
4C~29.30 is also consistent with this trend (Chandola et al. 2010).

In order to extend H{\sc i}-absorption investigations to a larger
number of GPS and CSS objects, especially those that are of lower
luminosity or more distant, and also towards the cores of larger
sources and the central regions of rejuvenated radio sources, we have
been observing these sources both with the Arecibo 305-m telescope and
the Giant Metrewave Radio Telescope (GMRT). The results obtained in the
first phase of this study were reported by Gupta et al. (2006).
Recently, the spectral region between 700 and 800 MHz became
temporarily available to radio astronomers due to the U.S.
television switch from analog to digital transmissions, with new
frequency allocations. We have used this opportunity to begin a
search for highly redshifted H{\sc i} and OH absorption within this
band against the continuum emission from  CSS/GPS radio sources of
appropriate redshift.  The first source observed, the well-known galaxy
CTA~21 (4C~16.09), shows strong H{\sc i} absorption.  We summarize
the properties of CTA~21 in Section 2, the observations and results
obtained with the Arecibo telescope in Section 3, and provide a
discussion and concluding remarks in Section 4.

\section{CTA~21 (4C~16.09)}
This ``classic'' radio source has been studied extensively. A deep
optical image of the field by Labiano et al. (2007) confirmed the
earlier identification of the source by Stanghellini et al. (1993) with
a faint galaxy having a $V$ magnitude of $\sim$23.4.  The optical
spectrum shows a weak continuum with bright [O{\sc ii}] and [O{\sc iii}]
lines, and yields a redshift of 0.907 based on five lines
(Labiano et al. 2007). The radio continuum spectrum of the source
shows a turn-over at approximately 1~GHz, below which the spectrum is
rather flat, or slightly inverted, down to at least 74~MHz (Steppe et
al.  1995; Torniainen et al. 2007). There is no evidence for
significant radio variability of the source (Altschuler \& Wardle 1977;
Bondi et al.  1996;  Aller et al.  1985).  It is at most very weakly
polarized, this being $<$0.5\% at frequencies below 5 GHz, increasing
to a few per cent at 15 GHz (Perley 1982; Aller et al.  1985).

The high angular resolution radio structure of CTA~21 has been
extensively imaged  over the years.  The early observations by
Clarke et al. (1969) at 408 and 448 MHz, and Kellermann et al. (1971)
at 1670~MHz suggested the existence of structure on different scales
although the details are difficult to relate to features seen in later
images.  Wilkinson et al. (1979) imaged the source at 609~MHz and
suggested that a compact component smaller than $\sim$12 mas
contributes 4.6 Jy to the total flux density of 8.7 Jy, and is embedded
in a 2.2-Jy halo with a size of 30$\times$15 mas$^2$ at a position
angle (PA) of 160$^\circ$. In addition, 1.9 Jy is attributed to a
fully resolved component which is $>$30 mas in size.  Wilkinson et
al.  also consider the possibility that their ``2.2-Jy halo'' and their
fully resolved component may be a single partially resolved component.
Jones (1984) imaged the source at 4831~MHz and found it to be an
asymmetric double with a separation of 12 mas along a PA of
160$^\circ$. He also noted that the components are slightly resolved
perpendicular to the source axis.  A combined MERLIN and VLBI  image
of CTA~21 at 1663 MHz (Dallacasa et al.  1995) with an angular
resolution of 40 mas shows a compact component which could correspond
to the double structure seen by Jones (1984) with more diffuse extended
emission to the south extending up to $\sim$300 mas.  Some evidence for
this feature is seen in the 15-GHz image by Spencer et al.  (1989).
The compact component, as imaged with a resolution of $\sim$4 mas by
Dallacasa et al.,  shows structure lying along the source axis, as
well as emission orthogonal to this.  On this basis, these authors
suggest that it could be classified as a lobe.  A VLBA image at
15~GHz with $\sim$2.4 mas resolution shows a number of components
within a region of size $\sim$40 mas (Kellermann et al.  1998).
Comparing the images of Jones (1984) and Kellermann et al.  (1998),
which are closest in frequency and resolution, the southern, brighter
component of Jones is possibly the brightest feature seen in the image
of Kellermann et al., while the location and PA of the northern
component of Jones would be consistent with the northern components
seen by Kellermann et al.  lying $\sim$10 mas from the dominant
component.  The weaker features of Kellermann et al. are not visible
in the image of Jones (1984).  Dallacasa et al. (1995) see emission
extending farther south in their 1663-MHz image when compared with the
15-GHz image of Kellermann et al.  (1998).

\section{Observations and Results}

On 2009 June 12, and for a limited period, a largely unexplored region of
frequency space between 700 and 800 MHz  opened up to radio
astronomers when US TV stations were mandated to switch from analog to
digital transmissions, with new frequency allocations.  When aware of
the opportunity to make  observations in this temporarily freed-up
band, the Arecibo Observatory prepared a receiver to cover it.
This is used in conjunction with the Mock spectrometer (named after its
designer/builder, the late Jeff Mock).  Using this combination, we have
undertaken a search of 29 CSS/GPS sources whose redshifts bring their
$\lambda$21-cm H{\sc i} or OH main lines into the 700--800 MHz band. At
the time of the observations described here, a system temperature of
$\sim$110~K was obtained on ``cold sky''. With a sensitivity of
$\sim$9~K/Jy, this gave a System Equivalent Flux Density (SEFD) of
$\sim$12.5~Jy.

The observations were carried out using the Double Position Switching
(DPS) technique (Ghosh \& Salter 2002). Each ON/OFF position-switched
observation with 5-min observing phases  made on CTA~21, was
followed by an ON/OFF of similar duration on the strong,
angularly-nearby, continuum source, 3C79, which served as a band-pass
calibrator. For sources with strong continuum emission, this strategy
is needed to produce acceptable spectral baselines.  In addition,
apparent features seen in the spectra can often be recognized as
astronomical, rather than due to  radio frequency interference
(RFI), from a comparison of the spectra for the target and calibrator
source.  Data reduction was performed using the Arecibo IDL analysis
package written by Phil Perillat.

In initial observations for this project, we accumulated  35~min of
``on-target'' integration for the H{\sc i} line against CTA~21.  The
frequency of the band center was offset by 2~MHz from that expected for
the redshifted H{\sc i} line for CTA~21 since the Mock spectrometer is
an AC-coupled device, producing a strong, but narrow, spike at the
band center.  The channel width of the spectra were 21\,kHz,
representing a velocity ($v = cz$) resolution of 16.1~km~s$^{-1}$ at the
frequency of the target line.  Both orthogonal polarizations of the
 signal were recorded.

Individual ON/OFF scans on CTA~21 were processed to yield
(ON$-$OFF) spectra, and these were bandpass corrected using
corresponding spectra for 3C79. Each individual CTA~21 and 3C79
spectrum was inspected for quality, and any RFI present noted.  All
proved to be acceptable within the frequency range where the
CTA~21 H{\sc i} absorption was expected to lie, and the
bandpass-corrected spectra were then co-added and the orthogonal
polarizations combined to produce a final spectrum.  A strong H{\sc i}
absorption line was detected close to the expected redshift for
CTA~21.  A polynomial was fitted to the continuum spectrum of
CTA~21, with the frequency range of the H{\sc i} absorption masked
out.  This polynomial was then used to derive the fractional absorption
across the line, as plotted against redshift in Figure~1. The 
fractional absorption  has an rms noise of 0.00055.

\begin{figure}
\epsscale{1.0}
\plotone{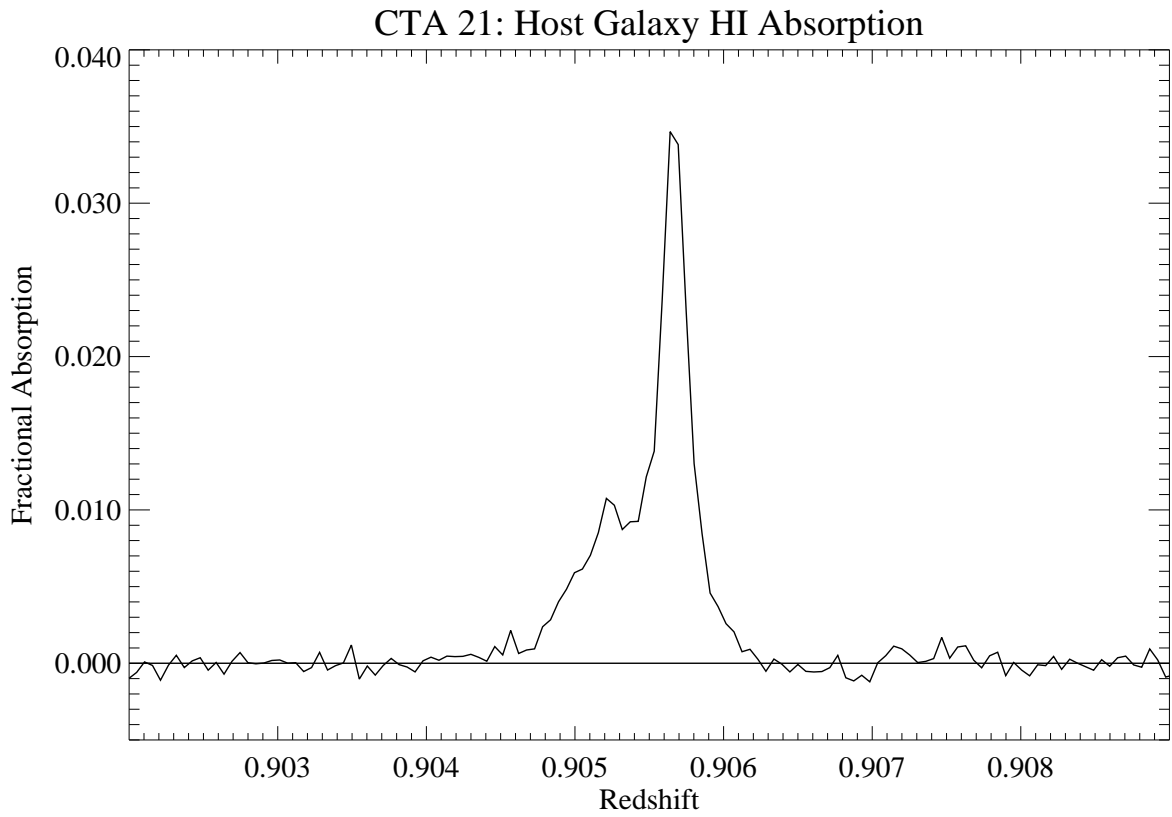}
\caption{The H{\sc i} 21-cm fractional-absorption spectrum
towards the GPS source CTA~21 plotted as a function of redshift.
The  fractional absorption has an rms noise of 0.00055.}
\end{figure}

The H{\sc i} line that we detect has a fractional depth of $0.034 \pm
0.001$. It is broad, with a width to half power of 67 km~s$^{-1}$, and
to zero intensity of $\approx$466 km~s$^{-1}$. The absorption spectrum,
expressed as optical depth, $\tau$, was integrated to provide an estimate of
the H{\sc i} column depth, {\it N}(H{\sc i}), for the region of the host
galaxy lying in front of the quasar continuum emission. This used;
\begin{eqnarray} {\it N}{\rm (HI)} & = & 1.835\times 10^{18}\frac{{\rm
{\it T}_{\rm s}}\int\!\tau (v)\,dv}{f_c} ~{\rm cm}^{-2} \label{eqcol}
\end{eqnarray}
where {\it T}$_{\rm s}$ is the H{\sc i} spin temperature (in K), $f_c$ is the
fraction of the background emission covered by the absorber, and $v$ is
the radial velocity ($v = cz$) in km~s$^{-1}$.  Assuming {\it T}$_{\rm s}$=100
K and $f_c$=1.0, a value of {\it N}(H{\sc i})$= 7.92 \times
10^{20}$~cm$^{-2}$ results.

The CTA~21 absorption line was also fitted by multiple Gaussians to
determine the peak optical depths ($\tau_p$) and FWHMs ($\Delta v$ in
km~s$^{-1}$) of likely individual components within the spectrum. The
best fit was obtained using 4 Gaussian components, as displayed in
Figure~2.  The parameters derived from the fits are given in Table~1,
and are discussed in Section~\ref{discussion}.  Here the
H{\sc i} column density for each component has been determined using;
\begin{eqnarray}
{\it N}{\rm (HI)} & = &
1.93\times10^{18}\frac{{\it T}_{\rm s}\tau_p\Delta v}{f_c}~{\rm cm}^{-2}
\label{eqcol1}
\end{eqnarray}
where {\it T}$_{\rm s}$, $f_c$ and $v$ are as above,
and we again assume {\it T}$_{\rm s}$=100 K and $f_c$=1.0.

\begin{figure}
\epsscale{1.0}
\plotone{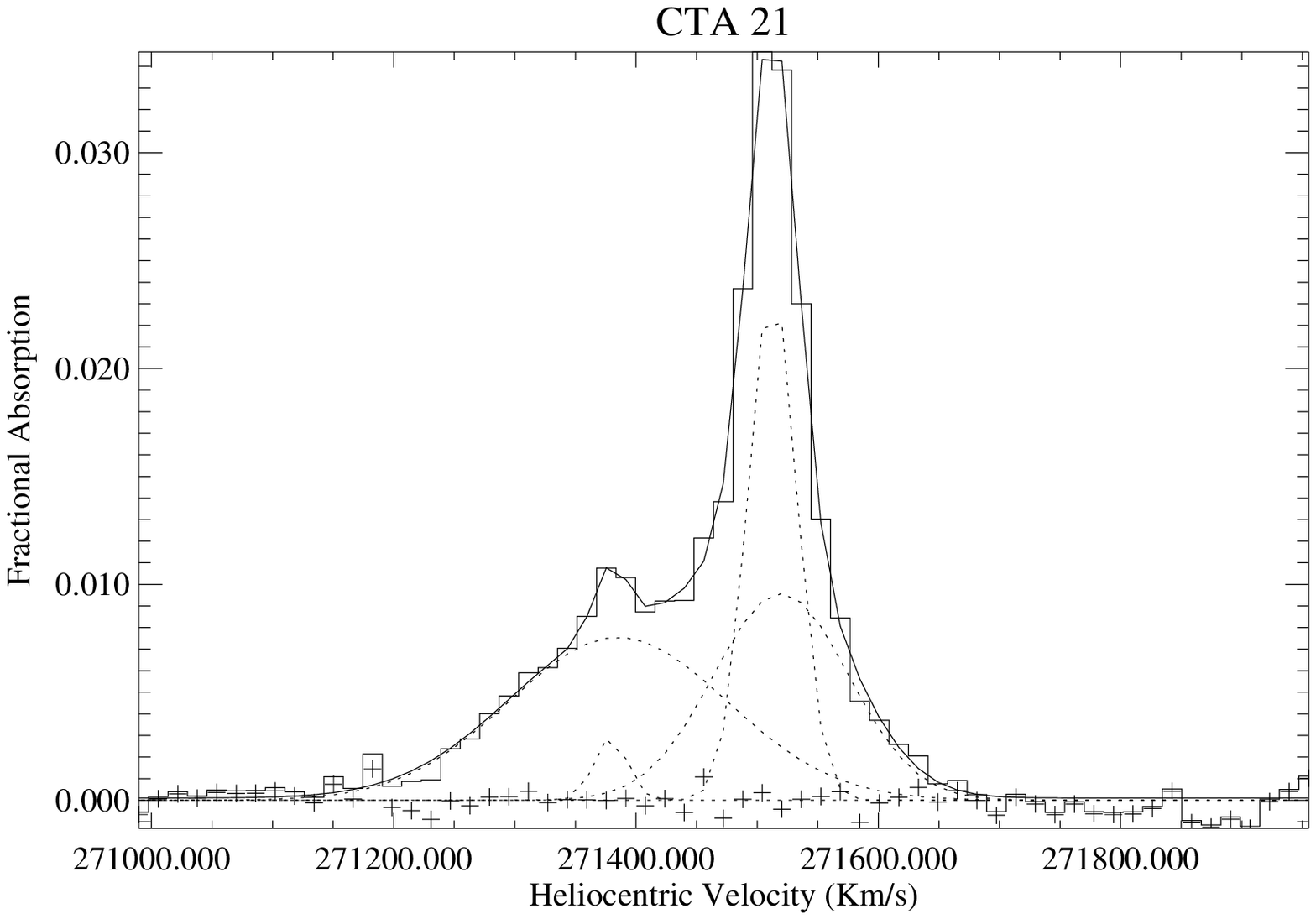}
\caption{The best-fit Gaussian decomposition of the H{\sc i} 21-cm
fractional-absorption spectrum towards CTA~21 as a function of radial
velocity. The four Gaussian components of the fit are plotted as dotted
lines, while plus signs show the residual spectrum after subtraction
of these components.}
\end{figure}

\begin{table}
\caption{Multiple-Gaussian fits to the H{\sc i} absorption spectrum of
CTA~21.}
\begin{center}
\begin{tabular}{ l l c l c c }
\hline
Id. & v$_{\rm{hel}}$ &   z  & FWHM & Frac. abs. & {\it N}(H{\sc i})               \\
no. & \kms           &      & \kms &            &  10$^{20}$cm$^{-2}$  \\
\hline
1 & 271379(9.7) &  0.90522 & 31.8(20.0) & 0.0029 & 0.018  \\
2 & 271385(79)  &  0.90524 &    212(77) & 0.0075 & 3.069  \\
3 & 271513(0.3) &  0.90567 &  47.6(0.8) & 0.0238 & 2.186  \\
4 & 271520(16)  &  0.90569 &    131(32) & 0.0096 & 2.427  \\
\hline
\end{tabular}
\end{center}
\label{cta21fit}
\end{table}

\section{Discussion and Concluding Remarks}
\label{discussion}
With a total column density of {\it N}(H{\sc i})$= 7.92 \times
10^{20}$~cm$^{-2}$, CTA~21 is a source with a high H{\sc i} column
density.  In the compilation of H{\sc i} absorption observations
towards CSS and GPS sources by Gupta et al. (2006), there are 96
sources in their `full sample', of which only five (J0111+3906,
J1357+4354, J1415+1320, J1819-6345, J1945+7055) have measured column
densities which are higher. It is also the highest redshift CSS or
GPS source for which H{\sc i} absorption has been detected. With an
overall linear size of $\sim$0.3 kpc, it is close to the upper envelope
of the {\it N}(H{\sc i}) vs projected linear size diagram (Pihlstr\"om et
al.  2003; Vermeulen et al. 2003; Gupta et al. 2006), and consistent
with the overall trend. The redshift of the strongest absorption
component in Table~1, $z \sim 0.9057$, agrees well with the
optically-determined value of 0.907 (Labiano et al. 2007).

The highest-resolution VLBI images at $\sim$5 and 15 GHz (Jones 1984;
Kellermann et al. 1998), with angular resolutions of a few
milliarcsec, do not show evidence for a core component. A
comparison of these two images suggests that the prominent components
have steep spectral indices. The estimated upper limit on the core
flux density of $\sim$10 mJy at 15 GHz implies that the fraction of
emission from the core is less than $\sim$1.5\% at this frequency,
which corresponds to an emitted frequency of $\sim$29 GHz.  Although
the {\it N}(H{\sc i}) vs core fraction diagram to test the unified scheme for
active galaxies and probe the geometry of the H{\sc i} disk has a large
scatter (e.g. Gupta \& Saikia 2006), CTA~21 appears broadly consistent
with this.  The non-detection of a radio core is also consistent with
its identification as a radio galaxy.

The existence of a compact double-lobed structure of size  $\sim$12
mas seen in the highest-resolution VLBI images (Jones 1984;
Kellermann et al. 1998), plus the more extended diffuse emission
discussed above, raises the possibility that CTA~21 may be
undergoing repeated cycles of activity. For example, the image of
Kellermann et al.  (1998) shows evidence of weaker emission on
opposite sides separated by $\sim$40 mas, while Dallacasa et al.
(1995) find evidence of diffuse emission extending up to $\sim$300 mas
towards the south.  The different resolutions of these images make it
difficult to reliably estimate spectral indices for the
different components.  However, on the basis of the compact
double-lobed structure and diffuse extended emission we classify CTA~21
as a candidate rejuvenated radio galaxy. Rejuvenated radio sources
cover a large range of linear sizes (see Saikia \& Jamrozy 2009, for
a review), and it has been suggested that jet activity in compact
radio sources may be intermittent on time scales of
$\sim$10$^4$$-$10$^5$ yr (Reynolds \& Begelman 1997). Saikia et al. (2007)
and Chandola et al. (2010) explored a possible relationship between
rejuvenation of radio or jet activity and the detection of H{\sc i} in
absorption. Unfortunately the number of sources is still small because
most rejuvenated radio sources have weak radio emission in the
central or nuclear region. The well-known examples of rejuvenated radio
sources where H{\sc i} absorption has been detected are the giant radio
galaxy 3C236 which has a projected linear size of $\sim$4250 kpc and
exhibits evidence of star formation (Conway \& Schilizzi 2000), the
giant radio galaxy
J1247+6723 with a GPS core (Saikia et al. 2007), the misaligned DDRG
3C293 (Beswick et al. 2004) which also exhibits fast outflowing gas
blue-shifted by up to $\sim$1000 km s$^{-1}$ (Emonts et al. 2005),
the large southern radio galaxy Centaurus A (Sarma
et al. 2002; Morganti et al. 2008), and the rejuvenated radio galaxy
4C~29.30 (Chandola et al.  2010).  The archetypal radio galaxy 
Cygnus A, which has been shown to have two cycles of radio activity
from radio and X-ray observations (Steenbrugge, Blundell \& Duffy 2008;
Steenbrugge, Heywood \& Blundell 2010), also exhibits significant
nuclear H{\sc i} absorption (Conway 1999).  While the
sample size needs to be increased, the detection of absorbing H{\sc i}
gas in  rejuvenated galaxies appears to be even more frequent
than for CSS and GPS objects.  Considering the GPS objects listed by
Gupta et al. (2006), these have the highest H{\sc i} detection rate of
$\sim$50\%, and a median column density of $\sim$3$\times$10$^{20}$
cm$^{-2}$. In comparison, the rejuvenated radio galaxies discussed here
have column densities in the range of $\sim$8$-$50$\times$10$^{20}$
cm$^{-2}$, and  tend to exhibit complex multi-component absorption
profiles. The estimated column density of CTA~21 would be consistent
with the range for the other rejuvenated radio galaxies.

\section*{Acknowledgements}
We than an anonymous referee for his/her helpful comments.
The Arecibo Observatory is part of the National Astronomy and
Ionosphere Center, which is operated by Cornell University under a
cooperative agreement with the National Science Foundation.

\end{document}